\documentclass[aps,prl,reprint,superscriptaddress]{revtex4-1}

 \usepackage{amsmath,bm}
 \usepackage{mathrsfs}
 \usepackage{amsfonts}
 \usepackage{graphicx}
 \usepackage{setspace} %leaves captions single space in draft mode
 \usepackage{graphicx}
 \usepackage{epstopdf}
 \usepackage{dcolumn}
 \usepackage{amsmath}
 \usepackage{epsfig}
 \usepackage{indentfirst}
 \usepackage{psfrag}
 \usepackage{subfigure}
 \usepackage{amssymb}
 \usepackage{color}
 \usepackage{xcolor}
 \usepackage{units} % rz
 \usepackage{siunitx}

 \usepackage{graphicx}% Include figure files
 \usepackage{dcolumn}% Align table columns on decimal point
 \usepackage{bm}% bold math
 \usepackage{natbib}
\usepackage{physics}
\usepackage{dcolumn}% Align table columns on decimal point
\usepackage{bm}% bold mathhttps://www.overleaf.com/project/5af95d0f3775594d14d4a052
\usepackage{natbib}
\usepackage[backref=none,bookmarksnumbered=true,bookmarks=true,bookmarksopen=true,colorlinks=true,
citecolor=blue,linkcolor=blue,anchorcolor=green,urlcolor=blue,unicode=false]{hyperref}

\usepackage{ulem}[normalem] %Whatch out, places underlining in Journal references. Use \normalem before references to deactivate this 

\makeatletter
\newcommand\colorsout[1]{\bgroup \markoverwith{\textcolor{#1}{\rule[0.5ex]{2pt}{0.4pt}}}\ULon}

\makeatother
\begin{document}

\title{Pair excitations of a quantum spin on a proximitized superconductor}

\author{Stefano Trivini}
  \affiliation{CIC nanoGUNE-BRTA, 20018 Donostia-San Sebasti\'an, Spain}

\author{Jon Ortuzar}
  \affiliation{CIC nanoGUNE-BRTA, 20018 Donostia-San Sebasti\'an, Spain}

\author{Katerina Vaxevani}
  \affiliation{CIC nanoGUNE-BRTA, 20018 Donostia-San Sebasti\'an, Spain}

  \author{Jingchen Li}
\affiliation{CIC nanoGUNE-BRTA, 20018 Donostia-San Sebasti\'an, Spain}
\affiliation{School of Physics, Sun Yat-sen University, Guangzhou 510275, China}

\author{F. Sebastian Bergeret}
 \affiliation{Centro de F\'isica de Materiales (CFM-MPC) Centro Mixto CSIC-UPV/EHU, E-20018 Donostia-San Sebasti\'an,  Spain}
\affiliation{Donostia International Physics Center (DIPC), 20018 Donostia-San Sebastian, Spain}

\author{Miguel A. Cazalilla}
  \affiliation{Donostia International Physics Center (DIPC), 20018 Donostia-San Sebastian, Spain}
\affiliation{Ikerbasque, Basque Foundation for Science, 48013 Bilbao, Spain}

\author{Jose Ignacio Pascual}
  \affiliation{CIC nanoGUNE-BRTA, 20018 Donostia-San Sebasti\'an, Spain}
\affiliation{Ikerbasque, Basque Foundation for Science, 48013 Bilbao, Spain}

\begin{abstract}

A magnetic impurity interacting with a superconductor develops a rich excitation spectrum formed by superposition of quasiparticles and spin states, which appear as Yu-Shiba-Rusinov and spin-flip excitations in tunneling spectra. Here, we show that tunneling electrons can also excite a superconducting pair-breaking transition in the presence of magnetic impurities, which is hidden for electrons on bare superconductors.  Combining scanning tunneling spectroscopy with theoretical modeling, we map the excitation spectrum of a Fe-porphyrin molecule on the Au/V(100) proximitized surface into a manifold of many-body excitations and follow their behavior across a parity-changing transition.  Pair excitations emerge in the tunneling spectra as peaks outside the gap in the strong interaction regime, scaling with the pair correlation. Our results unravel the quantum nature of magnetic impurities on superconductors and prove that pair excitations are parity detectors for magnetic impurities.

\end{abstract}
\date{\today}
\maketitle

\begin{figure}[t]
        \begin{center}
			\includegraphics[width=0.99\columnwidth]{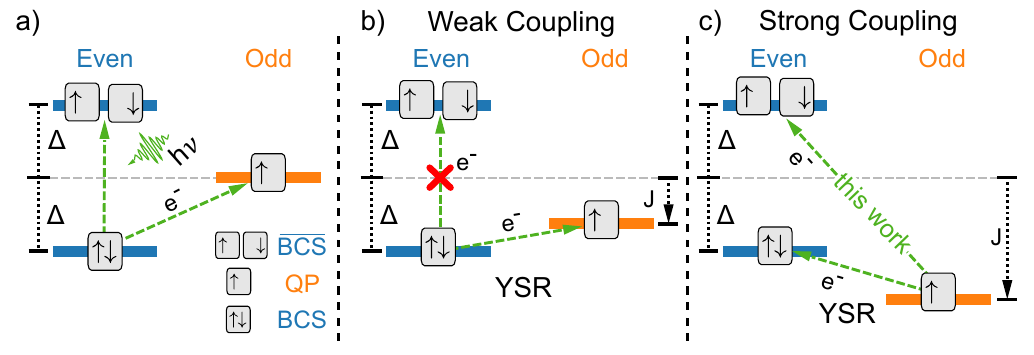}    
        \end{center}
			\caption{a) Scheme of the excitations of a superconductor with energy gap $\Delta$. The pair excitation ($\overline{\mbox{BCS}}$) can be probed by microwaves, while electrons can excite Bogoliubov quasiparticles (QP). The arrow boxes refer to the number of QP involved (eigenstates in the text). b) The exchange J induces YSR bound states below $\Delta$. Due to parity selection rule single electrons cannot excite the pair excitation ($\overline{\mbox{BCS}}$). c) Increasing J, the ground state becomes odd in parity, and the $\overline{\mbox{BCS}}$ state becomes accessible.}
		\label{fig1}
		\end{figure}

Superconducting materials provide an ideal platform for testing coherent dynamics of many-body states \cite{Janvier2015,park_steady_2022} and exploring their potential as  qubits \cite{Hays2021}. Pairing effects lead to a  many-body ground state formed by a condensate of Cooper pairs, protected from quasi-particle excitations by a pairing energy gap $\Delta$. Excitation of the superconducting ground state can be achieved by electrons \cite{cortesdel_rio_observation_2021,schneider_precursors_2022}, correlated pairs in Josephson currents \cite{van_den_brink_1991,Bretheau2013d}, or microwave photons \cite{glover_transmission_1956,Metzger2022,mannila_superconductor_2022}. In bulk superconductors, these excitations populate  a continuum of Bogoliubov quasiparticles (QPs) and admix with other states that quickly quench their quantum coherence. Sub-gap quasi-particle excitations, in contrast, can live long in a coherent state, allowing detection and  manipulation  of their quantum nature with   high fidelity. For example, sub-gap Andreev bound states in a  proximitized link  between two superconductors host addressable doublet quasiparticle  and  singlet pair-breaking excitations that can store quantum information  \cite{Janvier2015,Hays2021}. Population of these excited states follows parity-conserving rules: QP states are  odd in fermion parity and can be excited by adding or removing a fermion to the even-parity BCS ground state [Fig.~\ref{fig1}(a)]. 
Pair excitations involve the creation of \textit{two} correlated QPs into an excited state and, thus, have even-parity \cite{bardeen_theory_1957}. Therefore, they are accessible by absorption of one microwave photon or by addition of two particles with opposite spin.

Sub-gap excitations can be also produced by tunnelling electrons from a scanning tunneling microscope (STM). The most frequently observed case are  excitations of Yu-Shiba-Rusinov (YSR) \cite{Yu,Shiba,Rusinov} states, appearing when a magnetic impurity interacts with a superconductor via magnetic exchange $J$. YSR excitations appear in tunneling spectra as sub-gap bias-symmetric pairs of narrow peaks  \cite{Ji2008,Heinrich2018}.  In the regime of weak exchange interaction $J$ compared to the pairing energy $\Delta$, the peaks indicate the excitation of long-lifetime QP states obtained by adding a tunneling electron/hole into the BCS ground state. Pair excitations are, however, forbidden because this would require tunneling of two correlated electrons simultaneously [Fig.~\ref{fig1}(b)].

 \begin{figure}[th]
			\includegraphics[width=1\columnwidth]{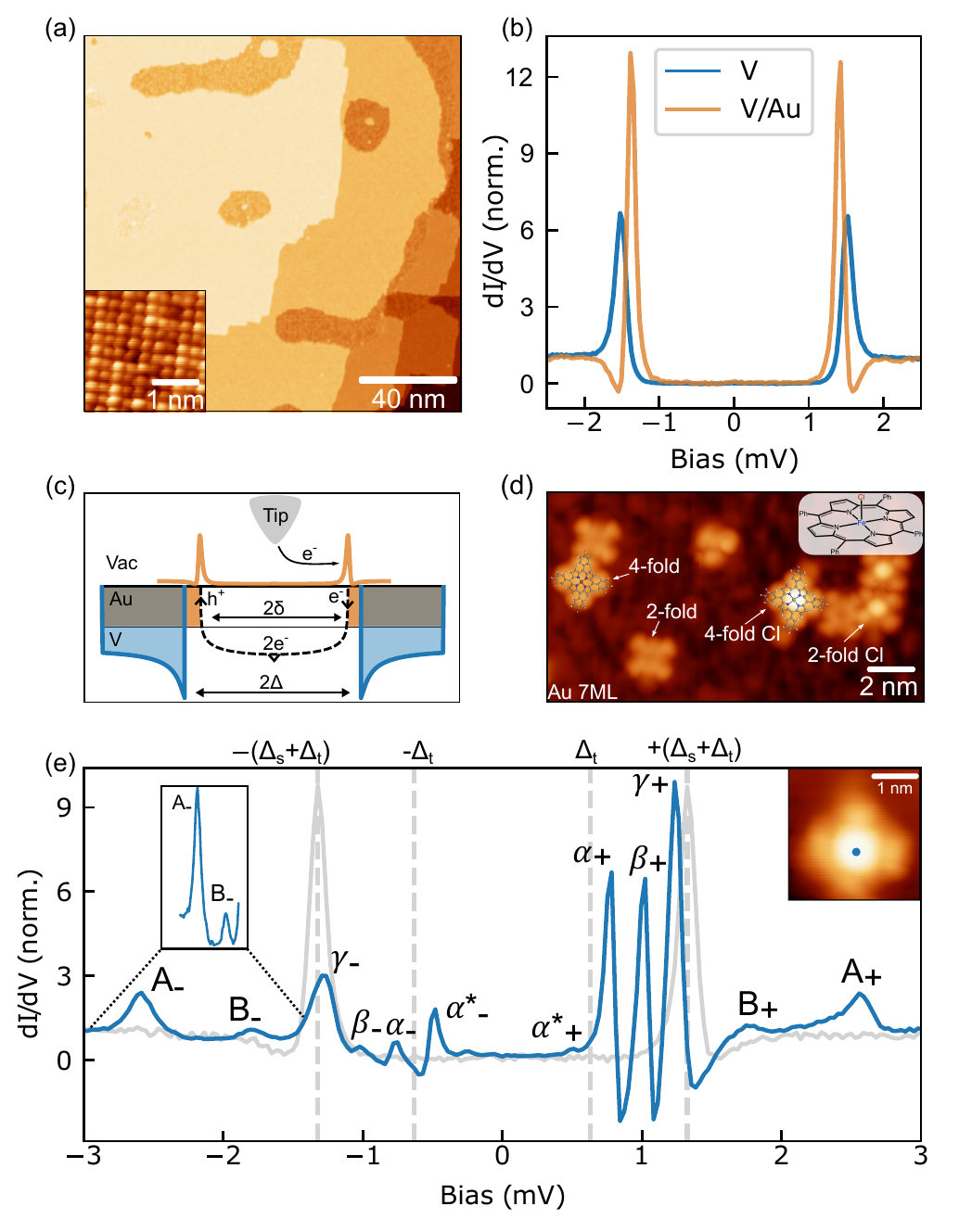}    
			\caption{a) STM image of the epitaxial film produced by depositing 2~ML of Au on V(100) and annealing to 550$^\circ$C ($\textrm{V}_S$ = 10 mV, \textrm{I} = 100 pA). Inset: constant height STM image of the film's squared lattice with 3 \r{A} unit cell ($\textrm{V}_S$ = 10 mV).
			b) dI/dV spectra measured on V(100) and on Au/V.   
			c) Scheme of the proximity effect; Andreev reflections at the interface with the V(100) deplete the film DOS and open a gap in the normal metal.  
			d) STM image with Cl functionalized tip, showing the different  FeTPP and FeTPPCl species on 7~ML Au/V(100) ($\textrm{V}_S$ = 300 mV, \textrm{I} = 30 pA), Inset: chemical structure of FeTPP-Cl. 
			e) \textit{dI/dV} spectrum measured on the center of a 4-fold FeTPPCl molecule, indicating the two extra-gap states (A,B) and the four intra-gap resonances ($\alpha,\alpha^*,\beta,\gamma$). In gray:  \textit{dI/dV} spectrum  on the bare substrate ($\textrm{V}_S$ = 3 mV, \textrm{I} = 75 pA). STM and STS data analysed with the WSxM \cite{Wsxm} and SpectraFox \cite{Spectrafox} software packages.}
		\label{fig2}
		\end{figure}

Here, we report the observation of pair-excited states in the YSR excitation spectrum of an Iron Porphyrin molecule on a proximitized gold thin film by means of scanning tunneling spectroscopy.   
Owing to the  magnetic anisotropy of the molecule, the YSR splits in multiple resonances both inside and outside the proximitized gap. Additionally, the molecules exhibit a new resonance amounting to twice the proximity induced gap $\Delta_s$ of the gold film, which we attribute to a pair excitation. 
Supported by  model calculations  for quantum spins, we show that inducing pair excitations with single particles  does not contradict parity-conserving rules because it is only observed when the magnetic molecule lies in a Kondo-screened regime. This regime is accessed when the exchange $J$ becomes larger than $\Delta_s$, and  
the magnetic impurity captures a QP, thus becoming Kondo-screened  \cite{Matsuura1977,Franke2011,Hatter2015,Farinacci2018,Malavolti2018}. From this ground state with odd fermion-parity, single-particle tunneling allows now YSR excitations into even states such as the BCS  state and its higher-lying excitations $\overline{\mbox{BCS}}$  [Fig.~\ref{fig1}c].

Our measurements were performed at 1.2 K using a STM (SPECS GmbH) under ultra high vacuum conditions. 
We used a V(100) single crystal as superconducting substrate (critical temperature T$_c$= 5~K and superconducting gap $\Delta_{\rm{V}}$(1K)=0.75~meV).  After sputtering and annealing to 1000$^\circ$C, the V(100) surface  appears with the characteristic V(100)(5x1) oxygen reconstruction \cite{KOLLER200111,DULOT2001172}, which does not affect the superconducting properties of the surface \cite{Etzkorn2018,Spin_ast,karan2021superconducting,huang_quantum_2020,huang_tunnelling_2020}.  
The  V(100)(5x1) surface was then covered with gold films, with thicknesses ranging from 1 to 10 ML, and shortly annealed  to $\sim$550$^{\circ}$C to produce epitaxial layers [Fig.~\ref{fig2}a]. 
The metallic films show a square lattice with an inter-atomic spacing of 2.9 \AA\ [inset Fig.~\ref{fig2}a] compatible with a non reconstructed Au(100) surface~\cite{hammer_surface_2014,liew_high_1990}, albeit some intermixing with the vanadium substrate is expected \cite{Huger2005a}.

The proximitization of the gold thin film was ascertained by comparing \textit{dI/dV} spectra over the films and over the bare V(100)(5x1) surface [Fig.~\ref{fig2}(b)].  To enhance the spectral resolution at the base temperature (1.2 K) of our STM, we used superconducting tips, obtained by deep tip indentations in the V(100) substrate. The spectra on the vanadium substrate show an absolute gap and two sharp peaks at  $\pm (\Delta_t + \Delta_V)/e = \pm$ 1.5 mV [Fig.~\ref{fig2}(b)], corresponding to the convolution of the superconducting density of states of  tip ($\Delta_t$) and sample ($\Delta_V/e = 0.75$~mV).  Spectra on the investigated gold films also exhibit a hard gap with similar width to the vanadium substrate \cite{Island2017}, but with a pair of very sharp resonances at slightly smaller bias of $\pm(\Delta_s+\Delta_t)/e = 1.4$ mV. These resonances,  first described by de Gennes and Saint James \cite{deGennes1963}, arise from Andreev reflection processes at the normal-superconducting (NS) interface \cite{arnold-theory,wolf_proximity_1982,kieselmann1987self,Truscott1999}, and behave as (Bogoliubov) quasiparticle excitation resonances of the  proximitized gold films [Fig.~\ref{fig2}(c)]. Interestingly, the de Gennes and Saint James (dGSJ) resonances shift to lower energy with increasing film thickness \cite{Vaxevani2022}, which is a useful knob for tuning the proximity gap $\Delta_{s}$ in the experiment. 

Next, we deposited the organometallic molecule iron tetraphenylporphyrin chloride (FeTPP-Cl) [inset of Fig.~\ref{fig2}(d)] on the proximitized gold films. This species hosts a Fe$^{3+}$ ion with a S=5/2 spin and a easy plane magnetic anisotropy \cite{heinrich_protection_2013,Heinrich2015}. STM images like in Fig.~\ref{fig2}(d) show that some of the molecules  maintain the Cl ligand on the surface. These intact FeTPP-Cl species appear with two different shapes in the images: species with two-fold symmetric FeTPP-Cl interact weakly with the substrate \cite{Vaxevani2022}, while the four-fold symmetric molecules that we investigate here behave as quantum impurities coupled to the superconducting substrate.

Spectra on the four-fold FeTPP-Cl molecules are characterized by a complex pattern of intra- and extra-gap resonances, as summarized in Fig.~\ref{fig2}(e).  We typically find three intra-gap pairs of peaks ($\alpha_{\pm}$, $\beta_{\pm}$ and $\gamma{\pm}$) in the region between $\pm \Delta_t$ and $\pm (\Delta_t + \Delta_s)$. These resonances are direct YSR excitations and appear with larger intensity at positive bias due to finite potential scattering components in the tunneling \cite{Farinacci2020,RubioVerdu2021}. Since the $\alpha_{\pm}$ resonance  appears close to $\pm\Delta_t$, its thermal excitations $\alpha^*_{\pm}$ can also be observed below $\pm\Delta_t$ in the spectra.  

In addition, \textit{dI/dV} spectra show fainter peaks [$A_{\pm}$ and $B_{\pm}$ in Fig.~\ref{fig2}e] above the proximitized gap. Peak $A$ can be associated with a spin-flip excitation of the molecular spin multiplet     \cite{heinrich_protection_2013,berggren_spin_2014,berggren_theory_2015}. We attribute it to a $M_s=\pm1/2 \to  M_s=\pm3/2$ transition because it lies at 1.3 meV above the dGSJ peaks, the expected magnetic anisotropy energy (2$D$), being $D$=0.65 meV the value of the axial anisotropy constant for this molecule \cite{heinrich_protection_2013,Kezilebieke2019,Vaxevani2022}.  
The origin of peak $B$ observed at $\sim$ 0.6 meV outside the gap cannot be directly connected with inelastic spin transitions. Instead, as we demonstrate in the following, peak $B$ corresponds to the excitation of a pair of QPs over the superconducting condensate.

\begin{figure}[t]
			\includegraphics[width=1\columnwidth]{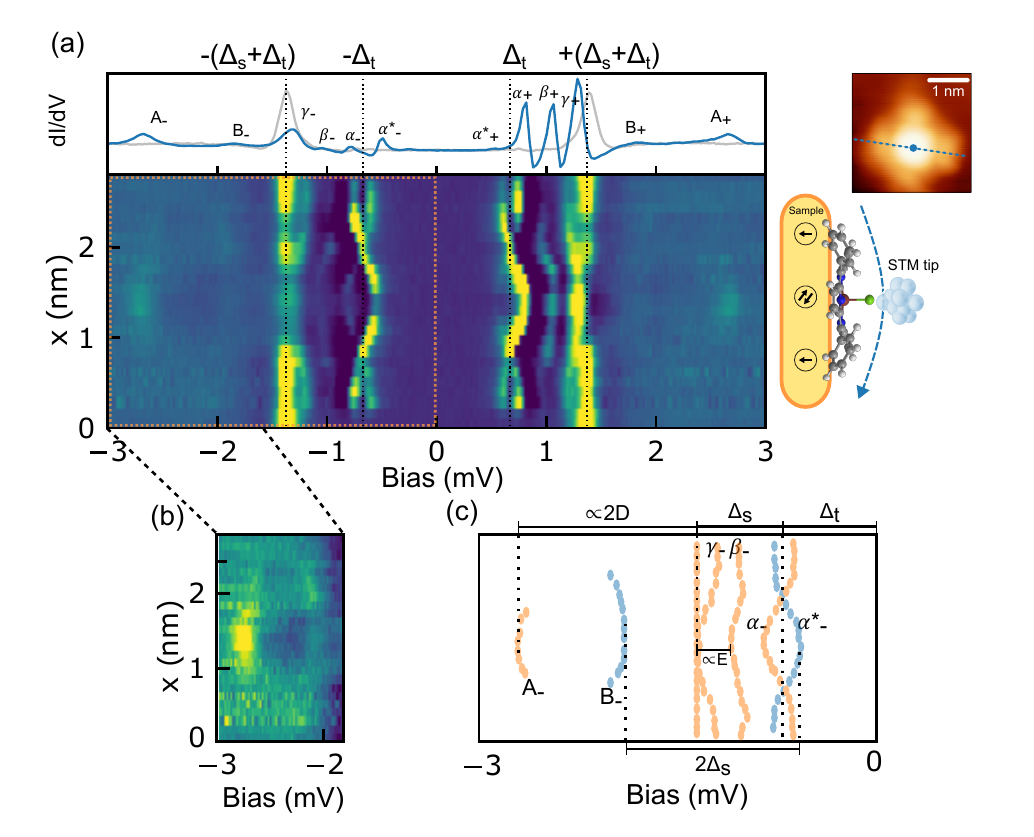}    
			\caption{a) Line of dI/dV spectra measured across a 4-fold FeTPPCl molecule (topography and sketch on the right) at constant current ($\textrm{V}_S$ = 3 mV, \textrm{I} = 75 pA). The single spectra on top is measured in the center and serves as a reference. b) Contrasted part of the out gap portion of the line-scan to highlight the signals A and B. c) Trace of the peaks energy positions along the line profile in the negative bias portion.}
		\label{fig3}
		\end{figure}

A hint on the origin of low energy excitations on superconductors can be obtained by measuring their evolution with variations of exchange coupling $J$  \cite{Franke2011,Hatter2015,Farinacci2018,Malavolti2018,chatzopoulos_spatially_2021}. We found that in standard tunneling regime the STM tip exerts an attractive force that distorts the \textit{flexible} molecular system and causes a controlled change of $J$.  Furthermore, this effect gradually decreases  as the tip is placed away from the center,
causing that \textit{dI/dV} peaks shift with the tip position along the molecular axis.
As shown in the spectral map of Fig.~\ref{fig3}(a), the three intra-gap YSR resonances shift to lower energies as the measuring position is laterally changed from the center of the molecule towards the phenyl groups. For the $\alpha$ state the shift is large enough to cross through the $\Delta_t$ line and exchange position with the thermal state $\alpha^*$. This is a fingerprint of a parity-changing quantum phase transition (QPT) in the ground state of the molecule-superconductor system. 

Unexpectedly, the extra-gap peaks $A$ and $B$  also change with $J$ [Fig.~\ref{fig3}(b)], but following a different trend: peak $A$ vanishes towards the sides, while the $B$ peak, which is much fainter in the center, move to higher energies at the molecule edges. The scheme of Fig.~\ref{fig3}(c) depicts the complex evolution of each intra-gap peak with the tip position, following the additional results presented in the supporting information (SI).
The apparent connection of the shifts of extra-gap peaks with intra-gap excitations suggests they are all related to the same many-body state, renormalized by changes in $J$ induced by tip. This state is formed by the spin S=5/2 of the quantum impurity, with $D\sim$0.65~meV, when it is coupled to the superconducting substrate with quasi-particle excitation peaks at $\Delta_{s}$.

\textbf{Theoretical model:} 
To interpret the results we used a minimal single-site model \cite{affleck_andreev_2000,vecino_josephson_2003}, extended for quantum impurities on superconductors by von Oppen and Franke \cite{von_oppen_yu-shiba-rusinov_2021,ysr_dimers_prb}. 
Calculations using this model are light and provide useful insights into the many-body spectrum of the system. The Hamiltonian reads:
\begin{equation}\label{H}
\begin{split}
    &H_s = H_0  + H_M + H_{J}  \\
    &H_0 = \Delta_s c^{\dagger}_{\uparrow}c^{\dagger}_{\downarrow} +\mathrm{h.c.} \\
    &H_M = DS_z^2 + E(S_x^2-S_y^2)\\
    &H_{J} = \sum_{\sigma\sigma'}c^{\dagger}_{\sigma}[J_zS_zs^z_{\sigma\sigma'}+J_{\perp}(S_+s^-_{\sigma\sigma'}+S_-s^+_{\sigma\sigma'})]c_{\sigma'}
\end{split}
\end{equation}
where $H_0$ describes a single-site superconductor, and $H_M$ accounts for the magnetic impurity spin anisotropy, also including transversal components \textit{E}. The term $H_{J}$ represents the (anisotropic) exchange coupling between the impurity and superconductor states, characterized by the exchange couplings $J_z$ and  $J_{\perp}$.  

\begin{figure}[t]
        \begin{center}
			\includegraphics[width=1\columnwidth]{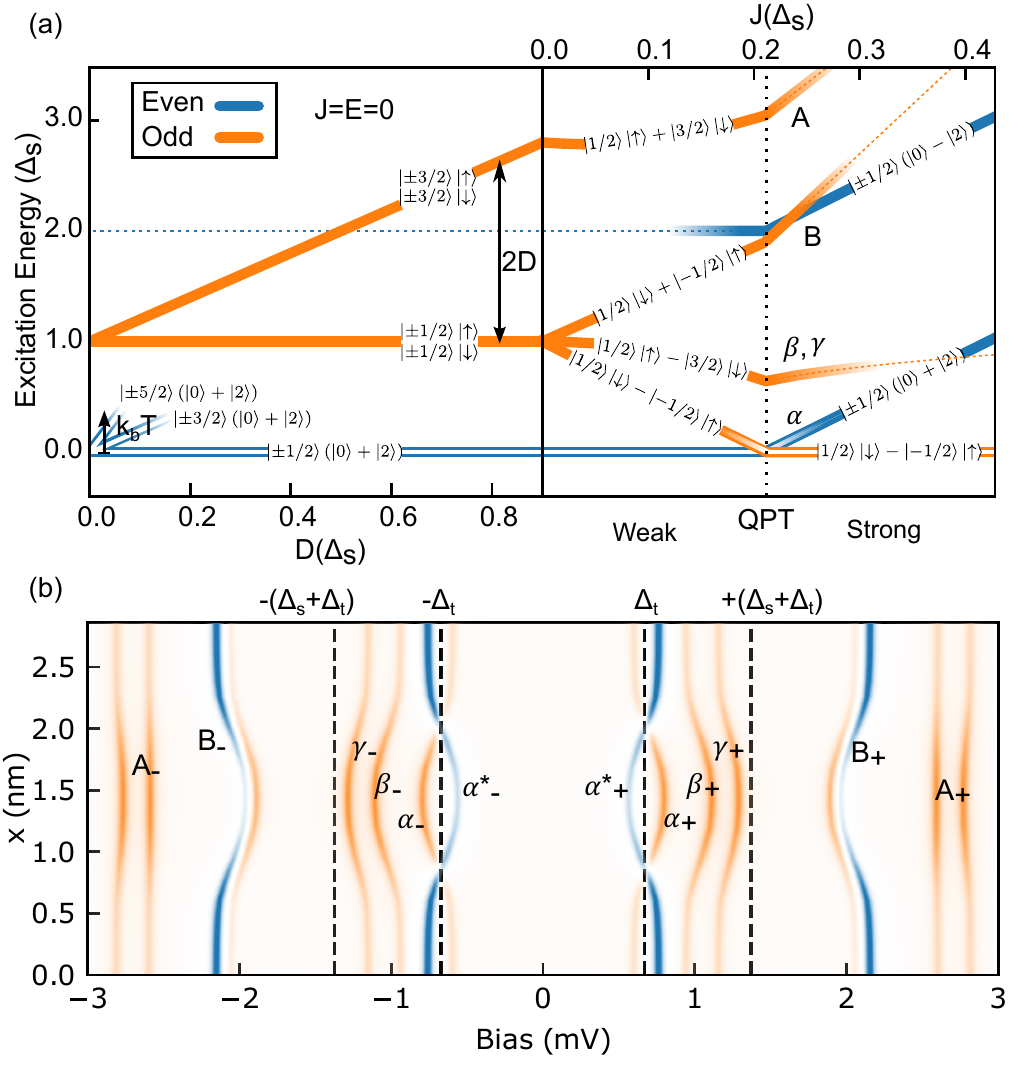}    
        \end{center}
			\caption{a) Ground state (empty lines) and permitted excitations for electrons (solid lines) obtained from the Hamiltonian of eq.~\ref{H}. In the left side $J$=$E$=0, the magnetic anisotropy $D$ splits the spin multiplets by 2$D$. In-gap states arise when the exchange coupling $J$ is turned on. Around the QPT, even and odd ground states are mixed by thermal excitations (vanishing lines). b) Simulation of a spectral map like in Fig.~\ref{fig3}a using the effective model, using as input the position of $\alpha$ and the re-normalized $D$ (\cite{SI}).}
		\label{fig4}
		\end{figure}

In Fig.~\ref{fig4}a we display the evolution of excitation energies in a tunneling experiment with increasing $D$ and $J$, obtained from the exact eigenstates of the  Hamiltonian in Eq.~\eqref{H}. Adding a tunneling electron (or hole) to the  ground state of the system leads to a change in fermion parity. Therefore, only transitions between even and odd parity states are allowed (blue and orange in Fig.~\ref{fig4}a). 
For the case of  negligible exchange $J$, the anisotropy $D$ of the molecule splits the spin multiplet into  non-degenerate levels of equal $S_z$ (left side in Fig.~\ref{fig4}(a)). The ground state is a product state of the molecular spin-doublet and  the superconductor BCS ground state:
\begin{equation}
    \ket{e}=\ket{\pm 1/2}\otimes\ket{BCS} = \ket{\pm 1/2}\otimes (\ket{\textrm{0}}+\ket{\uparrow \downarrow})\;
    \label{weak}
\end{equation}
Tunneling experiments in this regime resolve  peaks caused by a QP excitation at $\pm\Delta_s$, and by an additional spin excitation at $\pm(\Delta_s + 2D)$ \cite{heinrich_protection_2013,Kezilebieke2019,Vaxevani2022}. The  spin multiplet in the BCS ground state can also be thermally populated when $k_bT>2D$, as observed in Ref.~\cite{Hatter2015}.

A finite exchange coupling $J$ [right panel in Fig.~\ref{fig4}(a)] mixes the spin multiplet with QPs states into \textit{symmetric} and \textit{anti-symmetric}  combinations of entangled molecular-spin and  superconductor states with definite total spin projection $S^T_z$ \cite{Zitko2008,Zitko2011}. 
As shown in Fig.~\ref{fig4}(a), \textit{symmetric} states  appear as excitations outside the gap, while the \textit{anti-symmetric} ones correspond to intra-gap excitations. For example, the peak $A$ in our experiments corresponds to the excitation of the entangled symmetric state with $S^T_z$=1 \cite{Kezilebieke2019}, while the antisymmetric state is a sub-gap state split off from the YSR excitation in the presence of axial magnetic anisotropy  \cite{Zitko2008,Zitko2011,Hatter2015,von_oppen_yu-shiba-rusinov_2021}. In fact, by including a small transversal anisotropy  $E$  this state further splits into two, accounting for the resonances $\beta$ and $\gamma$ observed in the experiment (see supplementary information \cite{SI}).

\begin{figure}[t]
        \begin{center}
			\includegraphics[width=0.9\columnwidth]{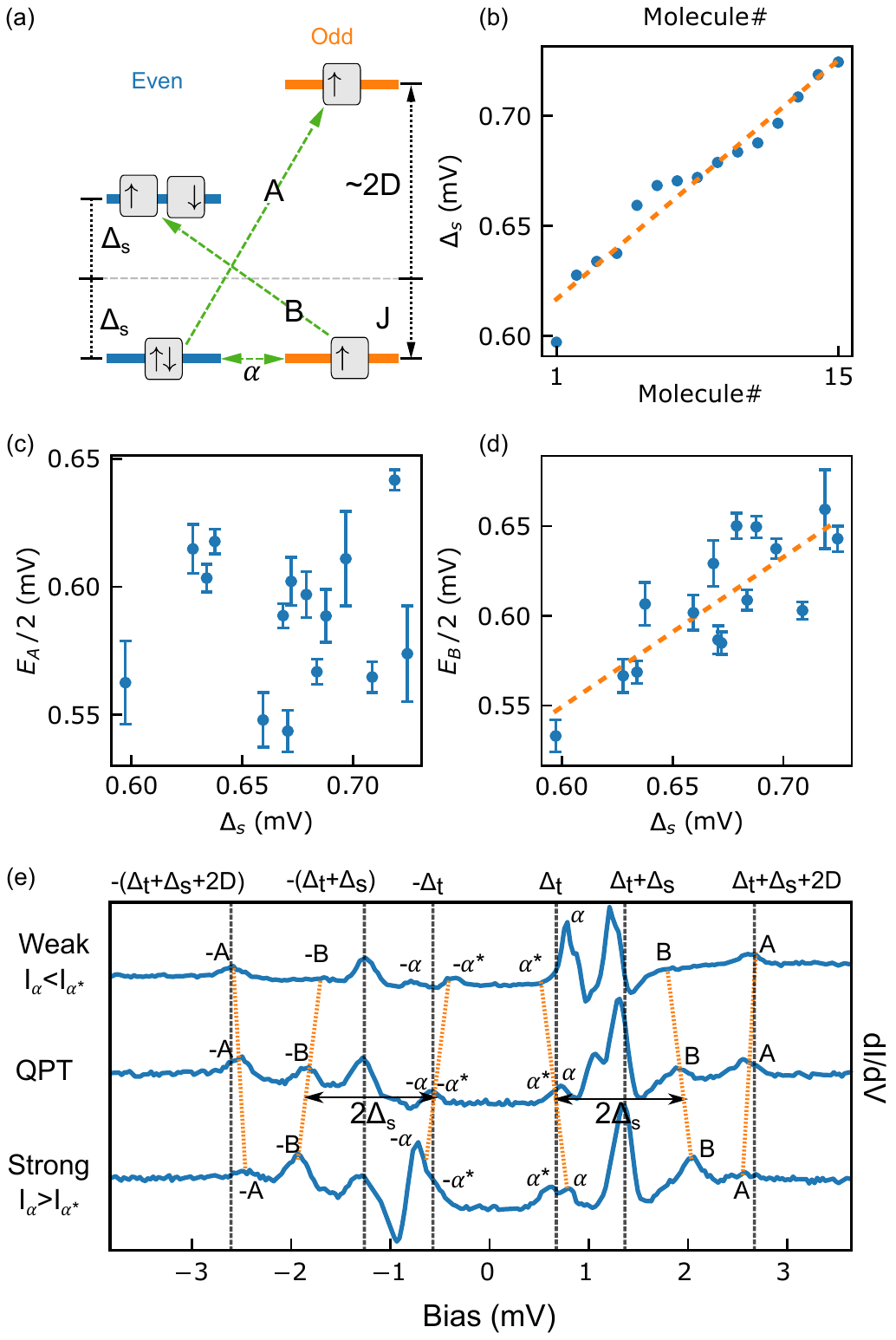}    
        \end{center}
			\caption{a) Scheme of $A$ and $B$ excitations ($J$ fixed to the QPT point); Peak $A$ scales with anisotropy $D$ and Peak $B$ with $\Delta_s$. b) Value of $\Delta_{s}$ measured close to 15 different molecules lying on different positions on the substrate, and on film with different thickness. c) Evolution of the position of peak $A$ for the 15 molecules, showing no correlation with $\Delta_{s}$. d) Evolution of the position of peak $B$ with $\Delta_{s}$, showing a linear dependence. e) Spectra of three molecules of the set, in weak, strong and at QPT (detected through their particle hole asymmetry). It shows the renormalization of the anisotropy, peak A shifts lower for higher $J$. Peak $A$ is more intense in the weak coupling case, while peak B with in strong coupling regime (further information in \cite{SI})}.
		\label{fig5}
		\end{figure}

Increasing $J$ above a critical value induces a QPT [Fig.~\ref{fig3}a],  where the ground state becomes an odd parity entangled state of  impurity's spin and a QP \cite{Balatsky,Farinacci2018}:
\begin{equation}
    \ket{o} = \ket{1/2}\ket{\downarrow}-\ket{-1/2}\ket{\uparrow}\; .
    \label{strong}
\end{equation}
From $\ket{o}$, only two even parity states can be reached by a tunneling electron or hole: the state \eqref{weak}, resulting in  YSR peaks $\alpha$, and the state:
\begin{equation}
    \overline{\ket{e}}=\ket{\pm1/2}\otimes\overline{\ket{BCS}}=\ket{\pm1/2}\otimes(\ket{0}-\ket{\uparrow \downarrow}).
    \label{pairExc}
\end{equation}
This second state is a pair excitation, i.e. the excitation of two QPs over the BCS state: $\gamma^{\dagger}_{\uparrow}\gamma^{\dagger}_{\downarrow}\ket{BCS}= {\overline{\ket{BCS}}}$ \cite{bardeen_theory_1957,SI}. The pair excitation lies at an energy $2\Delta_s$ above the YSR state and, as a consequence, the separation between the two even states is independent of molecular anisotropy [Fig.~\ref{fig5}(a)]. As we discuss next, peak B in the spectra corresponds to this pair excitation.

In Fig.~\ref{fig4}(b) we show a calculated spectral line profile simulating the experimental results  of Fig.~\ref{fig3}a, obtained by using  the model Hamiltonian of Eq.~\eqref{H}. The  input parameters for the calculation are simply the exchange coupling $J$, extracted from the position of $\alpha$,  the in-plane anisotropy $D=0.65$~mV, obtained from spin-excitation measurements on weakly coupled molecules \cite{Vaxevani2022}, and a fitted transversal anisotropy component \textit{E} \cite{SI}. The qualitative good agreement of theory results with the experimental observations confirm that the multiple  peaks can all be attributed to the excitation of  several manybody states formed by the impurity spin interacting with superconductor via a single orbital channel. 
Fermion parity selection rules  explain that peaks \textit{A}, $\beta$ and $\gamma$ fade away when the molecule enters in the strong interaction regime. Furthermore, the stronger intensity of peak B in this regime, and its shift with \textit{J} agrees with the behaviour of pair-excitated state $\overline{\mbox{BCS}}$.

To further corroborate the identification of peak $B$ as a pair excitation, we studied the evolution of peak A and B [Fig.~\ref{fig5}(a)] on $15$ molecules lying on different regions of the substrate and film thicknesses, with different measured values of $\Delta_s$ [Fig.~\ref{fig5}(b)]. 
In all these molecules, the position of peak $A$ with respect to $\Delta_s$, i.e. 2$D$,  is uncorrelated from  $\pm \Delta_{s}$ [Fig.~\ref{fig5}(c)]. On the other hand, the position of peak $B$, measured with respect to $\alpha$, scales with $\pm 2 \Delta_{s}$ [Fig.~\ref{fig5}(d)]. 

The different evolution of the extra gap peaks $A$ and $B$ with $J$ is shown in Fig.~\ref{fig5}(e), where we compare three spectra acquired on the center of three molecules with similar values of $\Delta_s$. The particle-hole asymmetry of the $\alpha$ YSR peak and its energy position  allows us to identify that they lie in the different interaction regimes indicated in the panel. Peak $A$ slightly shifts to lower energy with increasing $J$, due to renormalization of $D$ \cite{Kezilebieke2019}, and vanishes in the strong coupling case. Peak $B$, in contrast, becomes more intense in the strong coupling regime and follows the same the energy shift of $\alpha$, but spaced by $2\Delta_s$, as expected for the pair excitation.

\textbf{Discussion:} 
To date, pair excited states were only observed through adsorption of microwaves  \cite{Janvier2015,Hays2021} photons or Andreev pairs \cite{ Bretheau2013d}. Fermion-parity conservation forbids a single tunneling electron from  exciting a pair of Bogoliubov quasi-particles (the $\overline{\mbox{BCS}}$ state) in a superconductor. In our experiment, the  observation of the pair excitation with electrons was made possible by the existence of an odd-parity Kondo-screened ground state of a magnetic molecule  state on a superconductor, which enabled the excitation of two  even-parity states [Fig.~\ref{fig3}]: the BCS state, leading to the intra-gap  YSR resonance and the $\overline{\mbox{BCS}}$ pair excited state, observed as peak $B$. Even if this resonance appears outside the spectral gap, the pair state in the proximitized film  is a double population of a subgap state and, hence, it is expected to have a larger lifetime, facilitating its detection. 

It is also interesting to note that the quantum  spin model used here accounted for all observed resonances using just one single channel. Multiple sub-gap excitations resulted by entangled states of impurity and quasiparticles, mixed  by magnetic anisotropy constants $D$ and $E$. As we show in SI \cite{SI}, a small value of $E$ suffices to justify peak $\beta$, because the YSR excited state is integer and with large spin. This model successfully explains the important role of transversal and axial anisotropy and the effect of exchange on the magnetic anisotropy.

In conclusion,  we have used a proximitized gold film as a platform for studying many-body excitations in magnetic impurities \cite{Island2017}. The magnetic molecule FeTPP interacting with the substrate electrons host subgap YSR states and spin excitations outside the gap that are readily described by a a superposition of superconducting quasiparticles and impurity spin states using a zero-bandwidth model. Interestingly, we also found an excitation of a $\overline{\mbox{BCS}}$ pair state on molecules in the Kondo-screened regime,  which scales with the different pairing energy of proximitized films of different thicknesses. This is a hidden excitation of the BCS ground state for tunneling electrons that here was active for molecules with a bound quasiparticle. These results represent a novel route for addressing pair excitations on  proximitized superconductor.  Furthermore,  electron-induced pair excitations are a  smoking gun for unequivocally detecting the parity of the ground state \cite{huang_quantum_2020,Farinacci2018,Hatter2015}.

\begin{acknowledgments}
We acknowledge financial support from grants PID2019-107338RB-C61 and CEX2020-001038-M, PID2020-112811GB-I00) funded by MCIN/AEI/ 10.13039/501100011033, from the Diputación Foral de Guipuzcoa, and from the European Union (EU) through the Horizon 2020 FET-Open projects SPRING (No. 863098) and SUPERTED (No. 800923),  and the European Regional Development Fund (ERDF).
MAC has been supported by Ikerbasque, Basque Foundation for Science, and MCIN Grant No. PID2020-120614GB-I00 (ENACT).
F.S.B thanks Prof. Bj\"orn Trauzettel for his hospitality at W\"urzburg University,
and the A. v. Humboldt Foundation for financial support. 
MAC has been also supported by Ikerbasque, Basque Foundation for Science.
J.O. acknowledges the scholarship   PRE\_2021\_1\_0350   from the Basque Government.
\end{acknowledgments}

\end{document}

\title{Supplementary information:\\Pair excitations of a quantum spin on a proximitized superconductor} 

\author{Stefano Trivini}
  \affiliation{CIC nanoGUNE-BRTA, 20018 Donostia-San Sebasti\'an, Spain}

  \author{Jon Ortuzar}
  \affiliation{CIC nanoGUNE-BRTA, 20018 Donostia-San Sebasti\'an, Spain}

  \author{Katernina Vaxevani}
  \affiliation{CIC nanoGUNE-BRTA, 20018 Donostia-San Sebasti\'an, Spain}
 
\author{Jingchen Li}
\affiliation{CIC nanoGUNE-BRTA, 20018 Donostia-San Sebasti\'an, Spain}

\author{F. Sebastian Bergeret}
  \affiliation{Centro de Física de Materiales (CFM-MPC), 20018 Donostia-San Sebasti\'an, Spain}
\affiliation{Donostia International Physics Center (DIPC), 20018 Donostia-San Sebastian, Spain}

\author{Miguel A. Cazalilla}
  \affiliation{Donostia International Physics Center (DIPC), 20018 Donostia-San Sebastian, Spain}
\affiliation{Ikerbasque, Basque Foundation for Science, 48013 Bilbao, Spain}

\author{Jose Ignacio Pascual}
  \affiliation{CIC nanoGUNE-BRTA, 20018 Donostia-San Sebasti\'an, Spain}
\affiliation{Ikerbasque, Basque Foundation for Science, 48013 Bilbao, Spain}

\setstretch{1.1} 

\begin{abstract}
\vspace{2cm}
\tableofcontents
\end{abstract}
\date{\today}
\maketitle

\setstretch{1.0} 

\onecolumngrid\newpage
\section{Supplementary experimental results}
\twocolumngrid

\subsection{Line profile of a strongly coupled FeTPP-Cl}
Most of the FeTPP-Cl molecules investigated behave like the example shown in Fig. 3 of the main text. However, some species appear in a  different interaction regime.
In every case, we can identify the coupling regime by analyzing both the energy shift of the $\alpha$ and $B$ peaks with the tip position over the molecule and the particle-hole asymmetry (as shown in Fig. 5e). 

In Supplementary Fig.~\ref{LS_alt}  we study a FeTPP-Cl molecule in the strong coupling regime and show that this specie appears with protruding pair excitation spectral features. A spectrum on the center of the molecule (Supplementary Fig.~\ref{LS_alt}a) appears with  sub-gap YSR $\alpha$ peaks lying close to zero energy (detected at $\Delta_t/e$ bias, because of the superconducting tip), indicating that it lies very close to the quantum phase transition (QTP) between the two YSR interaction regimes. Accordingly, the  molecule shows both \textit{A} and \textit{B} peaks outside the superconducting gap with similar intensity.

    \begin{figure}[!b]
     \begin{center}
	       			\includegraphics[width=0.45\textwidth]{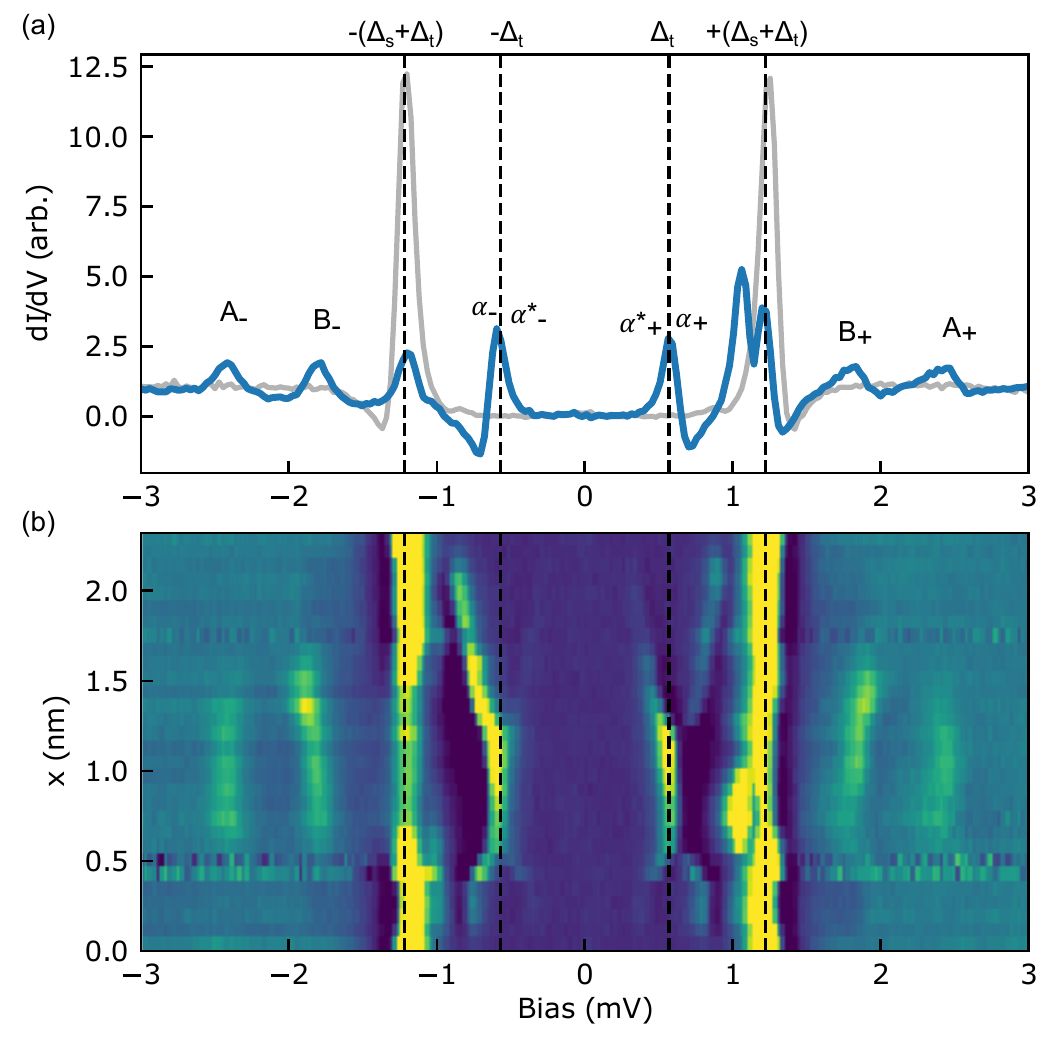}
	   \end{center}
			        \caption{a) Tunneling Spectrum of a $4$-fold FeTPP-Cl molecule measured at its center. The $\alpha$ peak is tuned to the quantum critical point of QPT whilst the $A$ and $B$ peaks outside of the gap are clearly distinguishable. b) For the same molecule, a line of $dI/dV$ spectra taken along the transversal direction. } \label{LS_alt}
		\end{figure}

As shown in  Supplementary Fig.~\ref{LS_alt}b, both the sub-gap YSR $\alpha$ peak and the pair excitation peak $B$ 
shift towards higher energies as the STM tip moves away from the molecule center. The stronger interaction regime  explains the strong pair excitation detected in the spectrum.

\subsection{Dependence of molecule-surface interaction on tip vertical position  }

The attractive effect induced by the STM tip over the molecule can be slightly controlled  via variations of tip proximity to the molecule. In Supplementary Fig.~\ref{LS_HDep}   we compare spectral profiles across the molecule in Fig.~3 of the manuscript with different junction resistances.  The increased interaction with the tip for the higher resistance case leads to larger variations of YSR peaks as the tip is moved across. 
This dependence with the tip's vertical distance also proves that the observed variations of $J$ are not an intrinsic property of the molecule but are induced by the interaction of the molecule with the STM tip \cite{chatzopoulos_spatially_2021}.

\begin{figure}[h] 
\begin{center}
 
    	    \includegraphics[width=0.49\textwidth]{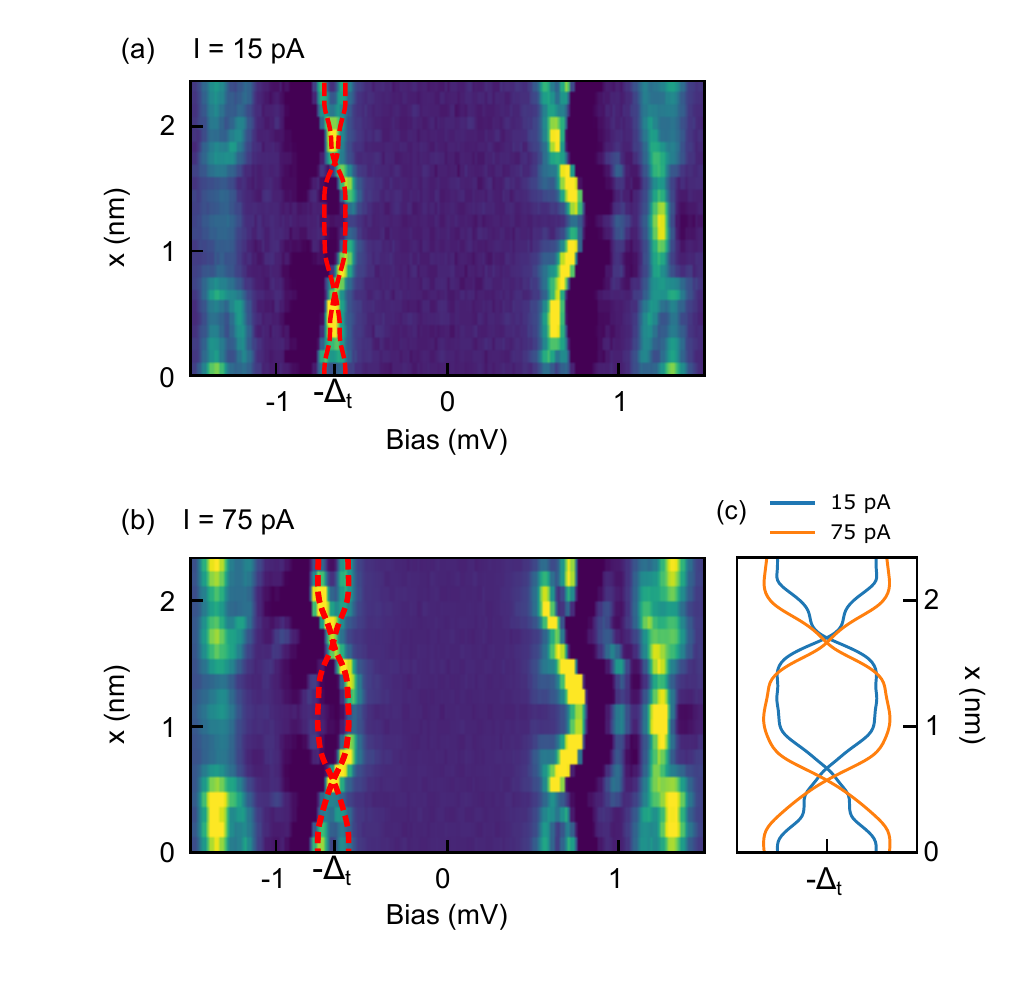}
    	    \end{center}
 	    	\caption{a,b) Linescan of the molecule presented in Fig.~3 of the main text at two set-point current values (V= 3mV I = 15, 75 pA). c) The $\alpha$ peak energy shift (dotted traces on the maps), both in strong and weak coupling, is enhanced by raising the tunneling current. }	\label{LS_HDep}

\end{figure}

\subsection{Dependence of pair excitation amplitude on the exchange interaction $J$}

In Fig.~5 of the manuscript we have shown that the pair excitation is a property of the molecule in the strong coupling regime (the Kondo-screened regime). In Supplementary Fig.~\ref{SFig3} we plot the amplitudes of the pair excitation peak $B$, measured at the center of a set of 15 FeTPP-Cl molecules, as a function of the corresponding exchange interaction $J$. The molecules are those studied in Fig. 5a of the manuscript. The value of J is obtained in each of them by fitting the position of YSR peaks $\alpha$ with the theoretical model. 

The intensity of the pair excitation $B$ is very small in the weak interaction regime (free-spin case). In this regime and at zero-temperature, this transition should be zero. The finite temperature of our experiment enables a small excitation probability due to the thermal population of the  $\alpha$ YSR state. The pair excitation amplitude increases significantly beyond the QPT, where now the ground state is odd in fermion parity and enables direct excitation of the pair state with a single tunneling electron. 
We fit the behaviour with a Boltzmann distribution obtaining a temperature T = $(1.2 \pm 0.2)$ K, that is compatible with the experimental temperature. 
\begin{figure}[h]
        \begin{center}
			\includegraphics[width=0.4\textwidth]{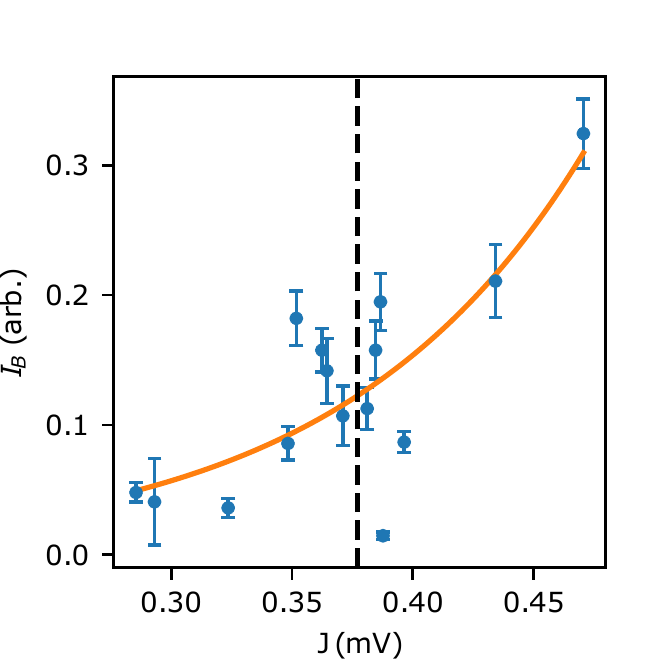}
        \end{center}
   
			\caption{Amplitude of the pair excitation as a function of the exchange interaction between molecule and surface $J$, for a set of 15 FeTPP-Cl molecules. The solid line is a Boltzmann fit }
		\label{SFig3}
	
\end{figure}
\vfill
\pagebreak 
\subsection{Spectrum of a FeTPP-Cl in the normal state}

\begin{figure}[h]
        \begin{center}
			\includegraphics[width=0.4\textwidth]{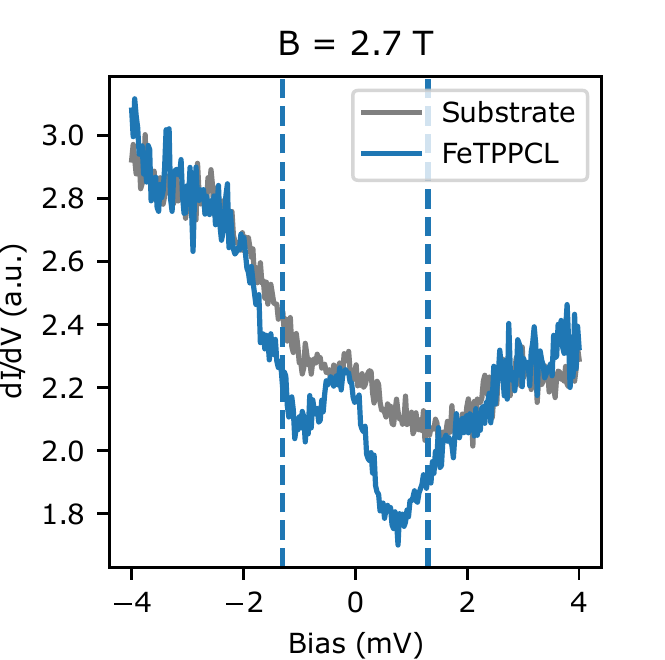}
        \end{center}
			\caption{dI/dV spectra of a 4-fold FeTPP-Cl with an applied magnetic field of 2.7 T applied in the direction perpendicular to the substrate. The superconducting gap of surface and tip is suppressed. }
			
		\label{grid}
\end{figure}

We quenched the superconducting state of tip and sample with a perpendicular magnetic field of 2.7 T to study the spectral shape of FeTPP-Cl  at low energy in the absence of superconductivity. The resulting spectrum (Supplementary Fig.~\ref{grid}) reflects the presence of Kondo-screening interactions and spin excitation, as we also find in the superconducting state. In particular, we observe a step excitation at ~1.3mV (dashed line in Supplementary Fig.~\ref{grid}) corresponding to the spin excitation $S_s=1/2$ to $S_z=3/2$ (D = 0.65 mV) and a weak Kondo resonance at 0 energy, signature of a ground state with $S_z=1/2$.

\clearpage

\onecolumngrid
\renewcommand{\thesection}{S\arabic{section}}
\section{Theoretical model}
\twocolumngrid

As mentioned in the main text, we describe the superconductor by using    a single-site model which is an extension of the one discussed in Ref.~\cite{von_oppen_yu-shiba-rusinov_2021}. Including the tip, the Hamiltonian reads:
\begin{equation}
    H_{\mathrm{model}} =H_s + H_t + H_{ts}, \label{eq:ham}
\end{equation}
where  $H_s$ and $H_t$ describe sample and tip, respectively, and $H_{ts}$ is the tunneling Hamiltonian;  $H_s$ is a single site superconductor coupled to a quantum impurity with  spin $S= 5/2$:
\begin{equation}\label{fmodel}
\begin{split}
    &H_s = H_0 + H_M + H_{J}  \\
    &H_0 = \Delta_s c^{\dagger}_{\uparrow}c^{\dagger}_{\downarrow} + \mathrm{h.c.} \\
    &H_M = DS_z^2 + E(S_x^2-S_y^2)\\
    &H_{J} =
    \sum_{\sigma\sigma^{\prime}} c^{\dag}_{\sigma} 
    \left[ V \delta_{\sigma\sigma^{\prime}} + J_z S_z
    s^z_{\sigma,\sigma^{\prime}} \right. \\
&\left. \qquad\qquad +
J_{\perp} \left( S_{+}s^{-}_{\sigma\sigma^{\prime}} + S_{-} s^{+}_{\sigma\sigma^{\prime}} \right) \right] c_{\sigma^{\prime}}.
 \end{split}
 \end{equation}
Here  $\Delta_s$ is the strength of the superconducting pairing  in the substrate,  $D$ and $E$ are the axial and transverse magnetic anisotropy of the molecule, and $J_z$ and $J_{\perp}$ are the axial and transverse magnetic exchange couplings,  and $V$ is the impurity  scattering potential. The effects of different terms on the spectrum of the system will be described in the following subsections. 

We also treat the Hamiltonian describing the superconducting tip, $H_t$, as a single-site superconductor:
\begin{equation}
H_t = \Delta_t c^{\dagger}_{t\uparrow}c^{\dagger}_{t\downarrow}+ \mathrm{h.c.}\; .
\end{equation}
Finally, the tunneling between the tip and sample is described by
\begin{equation}\label{Ht}
    H_{ts}=\sum_{\sigma}T_{\sigma\sigma'}c^{\dagger}_{t\sigma}c_{\sigma'}+ \mathrm{h.c.}\; .
\end{equation}
with $T_{\sigma\sigma'}=T_0+T_1\boldsymbol{S}\cdot\boldsymbol{\sigma}_{\sigma\sigma'}$, where $T_0$ is normal tunneling and $T_1$ is the spin-flip tunneling. Throughout we assume no Josephson current or multiple Andreev reflections between the tip and the sample, as expected in the weak tunneling regime. 

The Hamiltonian \eqref{eq:ham} is invariant  under time-reversal symmetry (TRS) and commutes with the  parity operator of combined the tip+sample system, $P_T=(-1)^{N_T}$, where $N_T = N_t + N_s$  is the total electron number operator. Notice that the tunneling Hamiltonian $H_{ts}$ does not commute with the sample (tip) parity operator $P_s=(-1)^{N_s}$ ($P_t = (-1)^{N_t}$). However, $P_s$ is still a good quantum number when considering the diagonalization of $H$ alone, as we shall do below. Thus,  the Hilbert space of the sample can be separated into two parity sectors: even parity with $P_s=1$ and odd parity with $P_s=-1$. In addition to TRS and parity, Eq.\eqref{eq:ham} exhibits other symmetries in certain  limiting cases. For example, in the limit where the transverse magnetic anisotropic $E$ vanishes, the Hamiltonian is invariant under the $Z_2$ symmetry that maps $S_{T,z}\to -S_{T,z}$
and interchanges $S_{T,x}\leftrightarrow S_{T,y}$, where $\boldsymbol{S}_{T} = \boldsymbol{S}+ 
\frac{1}{2} c^{\dag}_{\sigma} \boldsymbol{\sigma}_{\sigma\sigma^{\prime}}c_{
\sigma^{\prime}}$ is the total spin operator. This symmetry is generated 
by the rotation $ \hat{U} =  e^{i\pi S_{T,y}}e^{i\pi S_{T,z}/2}$. In addition,  we  neglect  the scattering potential $V$ in Eq.~\eqref{fmodel}. This potential  breaks particle-hole symmetry (PHS) and would modify the spectral weights of the peaks as 
mentioned in the main text.
However,  it does not modify the overall structure of the spectrum and only adds an additional fitting parameter to the model. Therefore, for the sake of simplicity, it can be taken to be zero, which renders the model invariant under PHS.

\subsection{Single site Superconductor}
To gain some basic understanding of the single-site superconductor model and fix the notations,  let us start by ignoring the magnetic molecule entirely:
\begin{equation}\label{H0}
    H_s= H_0=\Delta_s c^{\dagger}_{\uparrow}c^{\dagger}_{\downarrow}+\mathrm{h.c.}\; .
\end{equation}
The Hilbert space of the above Hamiltonian is a four dimensional linear space spanned by $\left\{\ket{2},\ket{0},\ket{\uparrow},\ket{\downarrow}\right\}$, where $\ket{\sigma =\uparrow,\downarrow} = c^{\dag}_{\sigma}|0\rangle$, $\ket{2}=\ket{\uparrow\downarrow}=c^{\dagger}_{\downarrow}c^{\dagger}_{\uparrow}\ket{0}$,  and $\ket{0}$ is the zero-particle or vacuum state.
In this basis, the Hamiltonian takes the following matrix form:
\begin{equation}
    H_0=\begin{pmatrix}0& \Delta_s& 0& 0\\
     \Delta_s &0& 0& 0\\
     0&0&0&0\\
     0&0&0&0\end{pmatrix}
\end{equation}
Upon diagonalization, the eigenstates are $\left\{|BCS\rangle=\frac{1}{\sqrt{2}}(\ket{2}+\ket{0}), |\overline{BCS}\rangle = \frac{1}{\sqrt{2}}(\ket{2}-\ket{0}),\ket{\uparrow},\ket{\downarrow}\right\}$ with (eigen-) energies $\{-\Delta_s,\Delta_s,0,0 \}$, respectively.  The eigenstates have well defined parity: $P_s|BCS\rangle  = |BCS\rangle$, $P_s|\overline{BCS}\rangle  = |\overline{BCS}\rangle$, i.e., they are even under parity, whilst
$P_s|\sigma\rangle = -|\sigma\rangle \, (\sigma=\uparrow,\downarrow)$, i.e. they are odd under parity.

Supplementary Fig. \ref{energy_diag} (a) shows the spectrum and the possible transitions between the different states of the sample in the single-site approximation. The even and odd parity states are connected by a single creation/annihilation operator of an electron (or  a Bogoliubov quasi-particle, see below). However, the two even-parity states are connected by two electrons/holes or two Bogoliubov operators, i.e. $\ket{\overline{BCS}}\propto\gamma^{\dagger}_{\uparrow}\gamma^{\dagger}_{\downarrow}\ket{BCS}$ (see Sec.~\ref{sec:pheno}). Note that the tunneling of a single electron from the tip will always change the parity of the sample.   
    
Let us consider now the tunneling between the tip and the sample at zero temperature, when  both,  the tip and the sample, are in their respective ground states:  $\ket{GS}=\frac{1}{2}(\ket{2}_t+\ket{0}_t)(\ket{2}_s+\ket{0}_s) = |BCS\rangle_t |BCS\rangle_s$. When e.g. a spin up electron tunnels from sample to tip (see Supplementary Fig. \ref{energy_diag} (a)), we have 
\begin{equation}\label{ts}
    c^{\dagger}_{t\uparrow}c_{\uparrow}\ket{GS}=\dfrac{1}{2}\ket{\uparrow}_t\ket{\downarrow}_s\; .
\end{equation}
This corresponds to a transition of energy $\Delta_s + \Delta_t$, whilst the  process in the opposite direction (from tip to sample) involves an energy $-(\Delta_t+\Delta_s)$.  This translates into two peaks at $\pm (\Delta_t + \Delta_s)$ in the tunneling spectrum, which mimic the coherence peaks observed in the tunneling between two s-wave superconductors.

\subsection{Spin-5/2 impurity with zero exchange coupling}
 
 Next, we consider how the excitation of the molecular  spin reflects on the tunneling spectra. Since $D$ is a large energy scale, we will first neglect the exchange coupling and set $J_{z} = J_{\perp}  =0$.  This limit is expected to capture some of the physics on the weak coupling side of the quantum phase transition (QPT, see below)~\cite{Kezilebieke2019,berggren_theory_2015}.  Now the Hamiltonian of the sample is $H_s = H_0 + H_M$, where the spin Hamiltonian $H_M$  accounts for the intrinsic magnetic anisotropy of the molecular spin:
\begin{equation}
    H_{M}=DS_z^2
\end{equation}
We assume easy-plane anisotropy ($D>0$) and, for the sake of simplicity, zero transverse magnetic anisotropy $E=0$. The effect of the latter will be discussed in the last subsection where the spectrum of full Hamiltonian is described.

 Since in this limit there is no exchange coupling, the Hilbert space of the sample is the tensor product
of the Hilbert space of the superconducting site and the molecular spin (see Supplementary Fig. \ref{energy_diag} (b)). We use the basis $\{(\ket{2},\ket{0},\ket{\uparrow},\ket{\downarrow})\ket{M}\}$, where $M$ is the eigenvalue of z-projection of the impurity spin, $S_z$. In the even parity sector  $S_{T,z} = S_z = M$, which is half-integer (recall that $S=5/2$) and therefore,
by TRS the eigenstates $\{\ket{BCS}\ket{M}$ ($\ket{\overline{BCS}}|M\rangle\}$) with the $S_{T,z} = \pm M$ are Kramers pairs and therefore degenerate in energy. In the odd parity sector, the $Z_2$ discussed above ensures the same for the eigenstates $\ket{\sigma}\ket{M}$.
 Since $D> 0$, the ground state is the doublet $S_{T,z}$ in the even parity sector, i.e. $\ket{GS}=\ket{BCS}\ket{\pm\frac{1}{2}}$. The eigenstates
 in the odd parity sector describe single (quasi-) particle
 excitations and, in this limit, have higher energy (see Supplementary Fig.~\ref{energy_diag}). 

 Let us  consider the tunneling of a single electron between the tip and the sample in this limit. The tunneling Hamiltonian, Eq.~\eqref{Ht}, contains a spin independent and spin-dependent terms with amplitude $T_0$ and $T_1$, respectively.  Since the tunneling current is second order in the tunneling amplitude,  there are three different contributions. The term   of order $|T_0|^2$ yields a spectrum identical to the one described in the previous subsection. The term of order $T_0T_1^*$ and its complex conjugate vanish due to TRS (but they would not in an external magnetic field that breaks TRS). Finally, the term of order $|T_1|^2$  accounts for the spin-flip processes which we discuss in the following. One of the possible tunneling processes is:
\begin{align}\label{tts}
    c^{\dagger}_{t\downarrow}c_{s\uparrow} S_+ \ket{GS}&=c^{\dagger}_{t\downarrow}c_{s\uparrow}S_+ \left[\ket{BCS}_t \ket{BCS}_s \ket{\tfrac{1}{2}} \right]\notag\\
    &\propto \ket{\downarrow}_t\ket{\downarrow}_s\ket{\tfrac{3}{2}}\; .
\end{align}
This process involves an excitation of the molecular spin and costs an energy $\pm(\Delta_s + \Delta_t +2D)$, the minus sign corresponding to  tunneling in the opposite direction (i.e. from tip to sample). Transitions (Supplementary Fig. \ref{energy_diag} (b)) to higher spin states are enabled by spin pumping \cite{heinrich_protection_2013}.

\begin{figure}[ht]
        \begin{center}
			\includegraphics[width=0.5\textwidth]{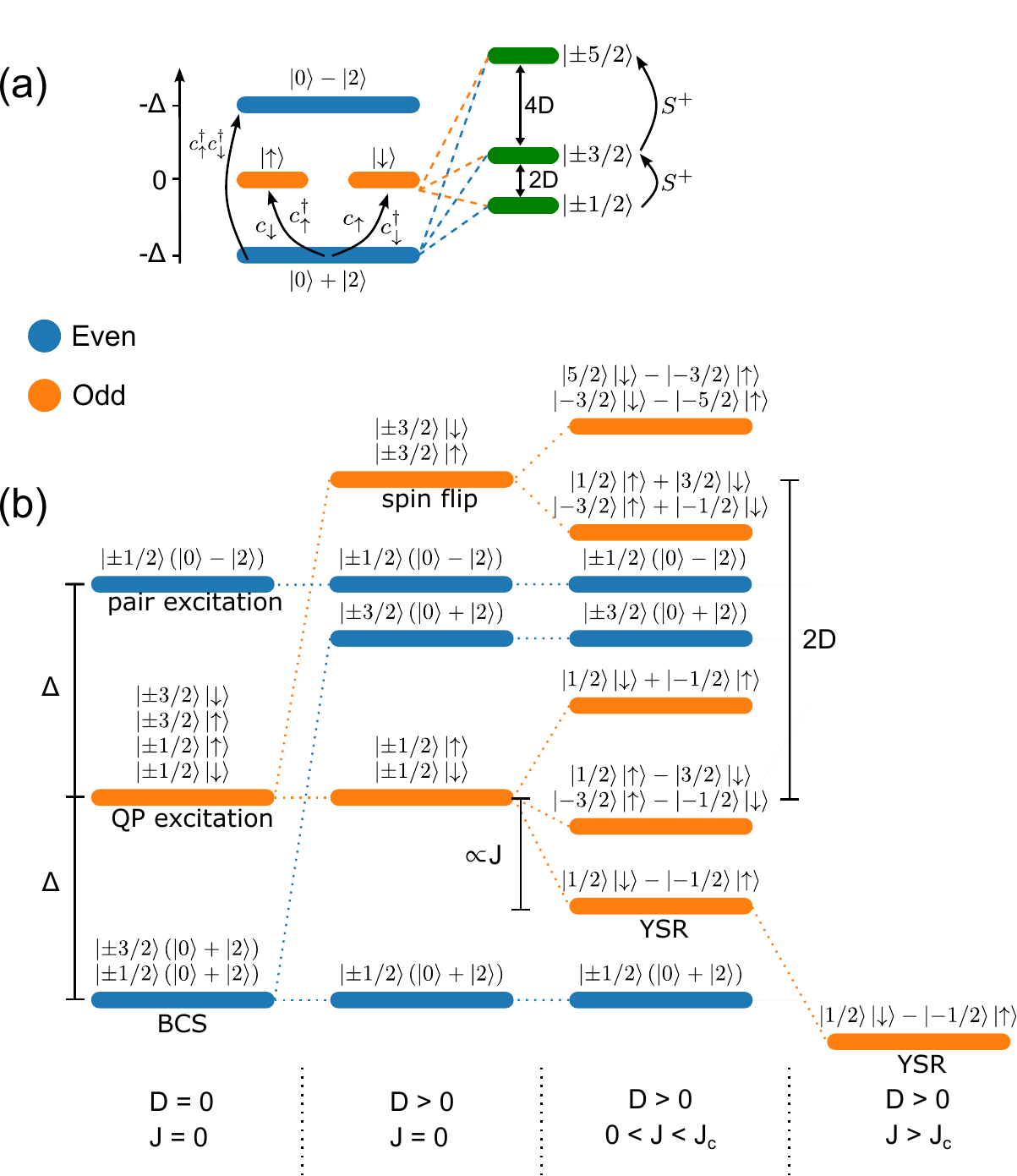}
        \end{center}
			\caption{(a) Spectrum of a single-site  superconductor with states labelled by fermion parity: even  (blue) and odd (orange) parity. When adding a spin $S = 5/2$  quantum impurity with easy-axis magnetic anisotropy $D$ but negligible exchange each eigenvalue split in three components. (b) Effect of a finite exchange coupling together with axial magnetic anisotropy on eigenvalues and eigenvectors. The parity changing QPT occurs for exchange grater than a critical value. To simplify the notation the coefficients of the linear combinations have been suppressed.  }
		\label{energy_diag}
	\end{figure}

\subsection{Spin-5/2 impurity with finite exchange coupling}

 Next, we account for the exchange coupling between the impurity and the substrate in the single site approximation and explain how the parity changing QPT takes place. The sample Hamiltonian is given in  Eq.~\eqref{fmodel}, where $H_J$ is the exchange term. We begin by investigating the isotropic limit where $J_z=J_{\perp}=J$ and $D=E=0$, i.e. $H_M =0$. The situation is not quite realistic  but makes the discussion of the QPT particularly clear.   
 
  In the isotropic limit,  the total spin of superconductor plus impurity $\boldsymbol{S}_T$ is conserved. Therefore, the eigenstates are organized into multiplets of $0\otimes \frac{5}{2} = \frac{5}{2}$, for the even parity sector with $P_s=+1$, and $\frac{1}{2}\otimes \frac{5}{2} = 2\oplus 3$, for the odd parity sector with $P_s=-1$. In the latter sector, the lowest energy state belongs to the multiplet with the smallest total spin, i.e. $S_T = 2$. Note that, by introducing a new energy scale $J > 0$, the ground state  is no longer uniquely determined
by $\Delta_s$ (see Supplementary Fig. \ref{energy_diag} (c)). In particular, the parity eigenvalue $P_s$ of the ground state  can change from even to odd by tuning $J$, resulting in a QPT \cite{Balatsky,Farinacci2018}.  The transition takes place when the energies of the lowest energy states in the even and odd parity sectors cross as $J$ increases. For a $S  = 5/2$ quantum impurity in the isotropic exchange limit  the critical  value is $J_C=4\Delta/7$.

    Regarding the overall structure of the spectrum, in the  even parity sector, the spin of the single-site superconductor is zero and therefore the exchange 
   coupling has no effect. The eigenstates take the form $\{\ket{BCS}\ket{M}, \ket{\overline{BCS}}\ket{M} \}$, i.e, there are two eigenstates per impurity spin $S_z=M$ projection. The states with the same superconductor component are Kramers pairs for $S_z = \pm M$  and therefore degenerate in energy. 
  
  On the other hand, in the odd parity sector, the exchange coupling is effective and the total spin of the
  eigenstates is an integer,  as discussed above. In the multiplet with $S_T = 2$,  the eigenstate  $(\ket{\tfrac{1}{2}}\ket{\downarrow}-\ket{-\tfrac{1}{2}}\ket{\uparrow})/\sqrt{2}$  with zero $S_{T,z}$ eigenvalue  becomes the lowest energy state. Indeed, for $D > 0$  both multiplets of $\boldsymbol{S}_T$ split, resulting in the states with the smallest $S_{T,z}$ eigenvalue \emph{from both multiplets} having the smallest energy. The $Z_2$ symmetry implies that the eigenstates with the opposite $S_{T,z}$ eigenvalue are degenerate.

\subsection{Full Hamiltonian}

\begin{figure}[ht]
        \begin{center}
			\includegraphics[width=0.5\textwidth]{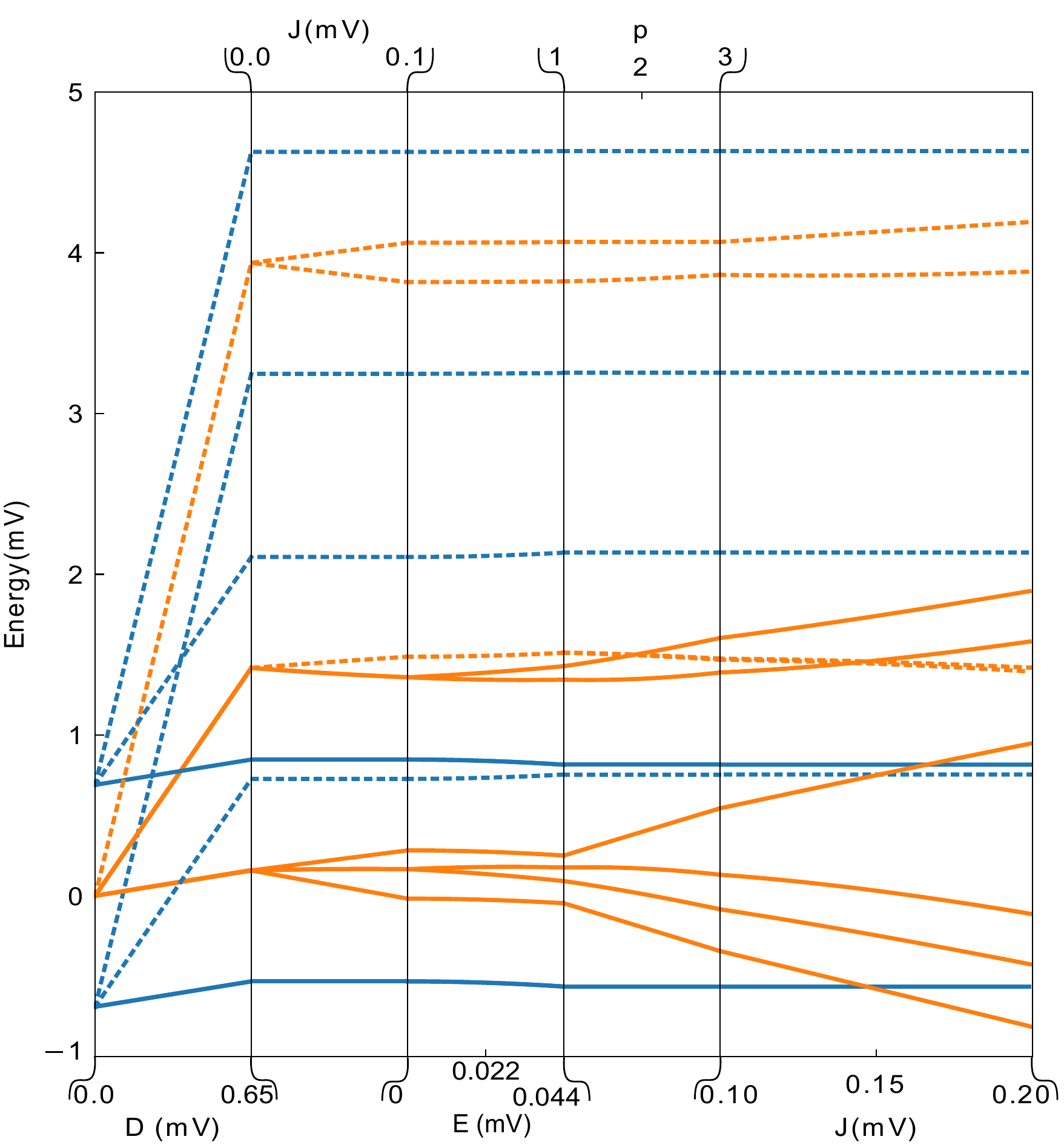}
        \end{center}
			\caption{Full evolution of even and odd parity eigenvalues of the Hamiltonian as a function of the model parameters: the magnetic anisotropy $D$, the exchange coupling $J$, transverse anisotropy $E$ and  the ratio of the transverse and longitudinal exchange $p$ couplings. The solid lines are the states reachable by tunneling one electron, while the dotted are prohibited by spin selection rules.}
		\label{energy_diag2}
	\end{figure}

 We now discuss the combined effect of all terms  in the model of Eq.~\eqref{fmodel}.
 In Supplementary Fig. ~\ref{energy_diag2} we show 
 the evolution of the spectrum  as the values of the different couplings are turned on up to values  compatible with the experimental ones.
The blue and orange lines correspond to the eigenstates with even and odd parity, respectively. The solid lines are the levels reachable by tunneling electron (at 0th order in $E$) resulting in the terms of order $|T_0|^2$ and $|T_1|^2$ in the tunneling current. 

 Consistent with the existence of large axial magnetic anisotropy $D$ in the molecular spin, we have assumed  an anisotropic exchange coupling where $J_{z}\neq J_{\perp}$. We find an optimal value for the ratio $p = J_{\perp}/J_{z} = 3$.   This value is close to the one obtained by projecting a $S=5/2$ spin onto the $S=1/2$ pseudo-spin describing the lowest energy doublet for a quantum impurity with $D > 0$~\cite{Zitko2008}. However, generally. the anisotropic exchange may  result from several different   mechanisms~\cite{Zitko2008}.

\begin{table*}[t]
\centering
\begin{tabular}{c||c|c|c}
     &Symmetries of the system& Nº of non-degenerate states.&
     Nº of non-degenerate states.\\
     & &Weak Coupling ($J>J_C$) &  Strong Coupling($J>J_C$)  \\
     \hline \hline
    $D=E=J=0$ &  TRS, $Z_2$, $\textrm{FSR}_{sub}$, $\textrm{FSR}_{mol}$  &2 &1 \\
    $D>0$, $E=J=0$ &  TRS, $Z_2$, $\textrm{FSR}_{sub}$ &6&3 \\
    $D,J>0$, $E=0$  &  TRS, $Z_2$&  6& 7\\
    $D,J,E>0$ &  TRS & 6 &12
\end{tabular}
\caption{Symmetries of the model in various limiting cases. The acronyms and symbols stand for TRS = time reversal symmetry, $\textrm{FSR}_{sub}$ = full spin rotation symmetry for substrate electrons, $\textrm{FSR}_{mol}$ = full spin rotation symmetry for molecular spin, $Z_2$= 180º rotation around the y-axis followed by a  90º rotation around the z-axis,  .}
\end{table*} \label{table}

Finally, we briefly discuss  the effect of the transverse magnetic anisotropy (E). In contrast to the anisotropic exchange, which does not break the $Z_2$ symmetry, the transverse anisotropy does. This has no effect on even parity states, where, due to the Kramers degeneracy, states with opposite total spin z-projection are degenerate. On the other hand, the breaking of $Z_2$-symmetry leads to splittings of the energy of odd parity states whose degeneracy is not protected by TRS. This effect accounts for the splitting seen in the $\gamma$ and $\beta$ peaks discussed in the main text.  As the value of $E$ used is rather small ($0.044$ meV), we  expect a small splitting. A first order perturbation theory calculation  for the degenerate states $\ket{\frac{1}{2},\uparrow}-\ket{\frac{3}{2},\downarrow}$ and $\ket{-\frac{1}{2},\downarrow}-\ket{-\frac{3}{2},\uparrow}$ yields a splitting of $\Delta E\sim 3\sqrt{2}E$. Thus,  we see that this small $E$ can account for a $\Delta E\sim 0.2$ meV separation between the two peaks.

The splitting of the remaining states is not substantial and can be ignored in a first approximation. Furthermore,  the admixture of the states  with different $S_{z,T}$ introduced by $E$ is very small and is not expected to introduce substantial changes to the discussion 
provided above.

\section{Phenomenology of the pair excitation}\label{sec:pheno}

Starting from the even-parity BCS ground state (weak coupling) in the zero-bandwidth model:
\begin{equation}
    \ket{BCS} = (u+vc^{\dagger}_{\uparrow}c^{\dagger}_{\downarrow})\ket{\textrm{vac}}
\end{equation}
From \cite{tinkham2004introduction} creation and annihilation operators of quasiparticles excitations are:
\begin{equation}
\begin{split}
        \gamma^{\dagger}_{\uparrow} &= uc^{\dagger}_{\uparrow}-vc_{\downarrow}\\
        \gamma^{\dagger}_{\downarrow} &= uc^{\dagger}_{\downarrow}+vc_{\uparrow}\\
        \gamma_{\uparrow} &= uc_{\uparrow}-vc^{\dagger}_{\downarrow}\\
        \gamma_{\downarrow} &= uc_{\downarrow}+vc^{\dagger}_{\uparrow}
\end{split}
\end{equation}
The odd-parity ground-state (strong coupling) is the BCS state with one excited quasiparticle:

\begin{equation}
\begin{split}
    \ket{0} &= \gamma^{\dagger}_{\uparrow}\ket{BCS}\\
    &=(u^2c^{\dagger}_{\uparrow}+uvc^{\dagger}_{\uparrow}c^{\dagger}_{\uparrow}c^{\dagger}_{\downarrow}-vuc_{\downarrow}-v^2c_{\downarrow}c^{\dagger}_{\uparrow}c^{\dagger}_{\downarrow})\ket{\textrm{vac}}\\
    &=(u^2+v^2)c^{\dagger}_{\uparrow})\ket{\textrm{vac}}\\
    &=c^{\dagger}_{\uparrow}\ket{\textrm{vac}}
\end{split}
\end{equation}
Where we drop 2 terms and commuted one using fermionic commutation relations.
From this ground state, by single electron tunneling, the system can return to the $\ket{BCS}$ state by annihilating the excited quasiparticle (YSR excitation):
\begin{equation}
    \gamma_{\uparrow}c^{\dagger}_{\uparrow}\ket{\textrm{vac}} = (u+vc^{\dagger}_{\uparrow}c^{\dagger}_{\downarrow})\ket{\textrm{vac}} =\ket{BCS}
\end{equation}
or exciting another quasiparticle and reaching the excited pair state:
\begin{equation}
    \gamma^{\dagger}_{\downarrow}c^{\dagger}_{\uparrow}\ket{\textrm{vac}} = (v-uc^{\dagger}_{\uparrow}c^{\dagger}_{\downarrow})\ket{\textrm{vac}} =\ket{\overline{BCS}} 
\end{equation}
 
 \vspace{5cm}

\bibliographystyle{apsrev4-1} 

%\bibliography{YSR-Bib} 
